\documentclass{aa}  
\usepackage{graphicx}
\usepackage{color}
\usepackage{graphicx}
\usepackage{lscape}
\usepackage{natbib}
\usepackage[varg]{txfonts}
\usepackage{url}
\usepackage{xspace}
\bibpunct{(}{)}{;}{a}{}{,}

\newcommand{\sao}{\object{SAO\,244567}\xspace}

\newcounter{Rco}
\newcommand{\Ionst}[1]{\setcounter{Rco}{#1}\Roman{Rco}}
\newcommand{\Ion}[2]{\mbox{#1\,{\scriptsize\Ionst{#2}}}}
\newcommand{\Ionw}[3]{\mbox{#1\,{\scriptsize\Ionst{#2}}~$\lambda\,#3$\,\AA}}

\newcommand{\Ionww}[3]{\mbox{#1\,{\scriptsize\Ionst{#2}}~$\lambda\lambda\,#3$\,\AA}}

\newcommand{\logg}{\mbox{$\log g$}\xspace}
\newcommand{\loggw}[1]{\mbox{$\log g\hspace{-0.5mm} =\hspace{-0.5mm}  #1$}}

\newcommand{\se}[1]{\mbox{Sect.\,\ref{#1}}}

\newcommand{\sla}{\raisebox{-0.10em}{$\stackrel{<}{{\mbox{\tiny $\sim$}}}$}}

\newcommand{\Teff}{\mbox{$T_\mathrm{eff}$}\xspace}
\newcommand{\Teffw}[1]{\mbox{$\Teff\hspace{-0.5mm} =\hspace{-0.5mm} #1 \,\mathrm{kK}$}}
\newcommand{\ebv}{$E_\mathrm{B-V}$\xspace}

\newcommand{\Lsol}{$L_\odot$}
\newcommand{\Msol}{$M_\odot$}

\newcommand{\Mdot}{$\dot{M}$}

\begin{document}

\title{The rapid evolution of the exciting star of the Stingray Nebula
           \thanks
           {Based on observations with the NASA/ESA Hubble Space Telescope, obtained at the Space Telescope Science
            Institute, which is operated by the Association of Universities for Research in Astronomy, Inc., under
            NASA contract NAS5-26666.
           }
           \thanks
           {Based on observations made with the NASA-CNES-CSA Far Ultraviolet Spectroscopic Explorer.
           }
       } 
\author{N\@. Reindl\inst{1}
        \and
        T\@. Rauch\inst{1}
        \and
        M\@. Parthasarathy\inst{2}
        \and
        K\@. Werner\inst{1}
        \and
        J\@. W\@. Kruk\inst{3}
        \and
        W.-R\@. Hamann\inst{4}
        \and
        A\@. Sander\inst{4}
        \and
        H\@. Todt\inst{4}}    

\institute{Institute for Astronomy and Astrophysics,
           Kepler Center for Astro and Particle Physics,
           Eberhard Karls University, 
           Sand 1,
           72076 T\"ubingen, 
           Germany,\\
           \email{reindl@astro.uni-tuebingen.de}
           \and  
           Inter-University Centre for Astronomy and Astrophysics,
           Post Bag 4, 
           Ganeshkhind,
           Pune 411007,
           India
           \and  
           NASA Goddard Space Flight Center, Greenbelt, MD\,20771, USA
	   \and
	   Institute for Physics and Astronomy, 
           University of Potsdam, 
           Karl-Liebknecht-Str\@. 24/25, 
           14476 Potsdam, 
           Germany}

\date{Received December 2013; accepted April 2014}

\abstract{\sao, the exciting star of the Stingray nebula, is rapidly evolving. Previous analyses suggested that it 
           has heated up from an effective temperature of about 21\,kK in 1971 to over 50\,kK in the 1990s.
          Canonical post-asymptotic giant branch evolution suggests a relatively high mass while previous analyses indicate
          a low-mass star.  
         }
         {A comprehensive model-atmosphere analysis of UV and optical spectra taken 
          during 1988\,--\,2013 should reveal the detailed temporal evolution of its
          atmospheric parameters and provide explanations for the unusually fast
          evolution.
         }
         {Fitting line profiles from static and expanding non-LTE
           model atmospheres to the observed spectra allowed us to study the temporal change
           of effective temperature, surface gravity, mass-loss rate, and terminal wind
           velocity. In addition, we determined the chemical composition of the atmosphere.
         }
         {We find that the central star has steadily increased its effective
           temperature from 38\,kK in 1988 to a peak value of 60\,kK in 2002. During the same
           time, the star was contracting, as concluded from an increase in surface gravity
           from \loggw{4.8} to 6.0 and a drop in luminosity. Simultaneously, the mass-loss rate
           declined from $\log$(\Mdot\,/\,\Msol\,yr$^{-1})=\,-9.0$ to $-11.6$ and the terminal wind velocity increased
           from $v_\infty = 1800$\,km/\,s to $2800$\,km/\,s. Since around 2002, the star stopped heating and has
           cooled down again to 55\,kK by 2006. It has a largely solar surface composition
           with the exception of slightly subsolar carbon, phosphorus, and sulfur. The results are
           discussed by considering different evolutionary scenarios. 
         }
         {The position of \sao in the log \Teff\ -- \logg\ plane places the star in the region of sdO stars. By comparison with 
          stellar-evolution calculations, 
          we confirm that \sao must be a low-mass star ($M < 0.55$\,\Msol). 
          However, the slow evolution of the respective stellar evolutionary models is in strong contrast to the observed fast
          evolution and the young planetary nebula with a kinematical age of only about 1000 years. 
          We speculate that the star could be a late He-shell flash object. Alternatively, it could be the outcome of 
          close-binary evolution. Then \sao would be a low-mass (0.354\,\Msol) helium prewhite dwarf after the common-envelope
          phase, during which the planetary nebula was ejected.
         }


\maketitle

\section{Introduction}
\label{sect:introduction} 

The long stellar evolutionary time scales mean it is in general
impossible for an astronomer to ``watch'' a star evolving in real time.
Intermediate-mass stars ($M_{\mathrm{ZAMS}}=0.8-8$\,\Msol) experience their most rapid evolution close
to the end of their nuclear-burning phase. Observing stars during
this period provides the unique opportunity to investigate on the
stellar asymptotic giant branch (AGB) and post-AGB evolution,
including the AGB mass-loss phase, and
the ejection, shaping, and excitation of planetary nebulae (PNe) --
phases that are still not fully understood.

\sao, the exciting star of the \object{Stingray Nebula} 
(\object{Henize 3$-$1357}, \citealt{henize1976}), is an unusually fast evolving star.
It was first classified to be a hot post-AGB star \citep{partha1989} based on the discovery of a 
circumstellar dust shell with far-IR (IRAS) colors and flux distribution similar to that
of PNe. Based on a spectral classification of the optical spectrum obtained in 1971, \cite{partha1995} concluded 
that the star was a B1 or B2 supergiant. From the UBV colors and the 1971 spectrum,
they estimated \Teffw{21}. However, they found that the optical spectra from 1990 and 1992 as well as the 
IUE\footnote{International Ultraviolet Explorer} spectra (1992 $-$ 1996) display many nebular emission lines, 
indicating that \sao has turned into a central star of a PN (CSPN) within a time span of only 20 years. 
Furthermore, when omparing the IUE spectra from 1988 and 1995 \cite{partha1995} discovered it is the only known CSPN 
that faded by a factor of 2.83 in its flux level within seven years. 
In addition, it was possible for them to observe how the stellar wind gradually decreased. From the IUE 
spectrum obtained in 1988, \cite{partha1995} measured a terminal wind velocity of $v_{\infty}=3500$\,km/s 
from the \ion{C}{IV} resonance doublet. The IUE spectrum in 1994 showed that the stellar wind vanished. 
Based on the IUE spectra, they estimated that \Teff\ must be around 55\,kK. 
Assuming a distance of 5.6\,kpc \citep{kozok1985} and an expansion velocity of 8\,km/s, 
\citet{partha1993} found that the post-AGB time of \sao is about 2700 years. 
Furthermore, they estimated the luminosity and core mass of the CS to be 
3000\,\Lsol\,\,and 0.55\,\Msol, respectively.

The first optically resolved images of the Stingray Nebula were presented by \citet{Bobrowsky1994} and 
\citet{Bobrowsky1998} using the Wide Field and Planetary Camera 1 (WFPC1) and WFPC2, respectively. 
\citet{Bobrowsky1994} found that in H\,$\beta$, the Stingray Nebula appears to have an equatorial ring of enhanced density 
tilted approximately 56$\degr$ from the line of sight. In addition, he found bubbles of gas above and below 
the ring with areas of decreased brightness near the poles where a fast stellar wind has broken through 
the red giant envelope. From the H\,$\beta$ flux, he derived an ionized mass of 0.2\,\Msol, $L=5000$\,\Lsol\, 
and a stellar core mass of 0.59\,\Msol. 
Thanks to the superior spatial resolution of the WFPC2 \citet{Bobrowsky1998} found evidence of collimated outflows, which 
are focused by the nebula bubbles and function like nozzles, with gas leaving through the polar holes. They also report a 
possible detection of a late type companion star at a distance of 2200\,AU from the central star. 

\citet{Umana2008} present the first detailed radio study of the Stingray Nebula by using the Australian 
Telescope Compact Array (ATCA). They find that the Stingray Nebula is still embedded in the dusty remnant of 
the AGB phase. Depending on their models, they derived an ionized mass of 0.057\,--\,0.07\,\Msol\,\,
and a total dust mass of 2$\times 10^{-4}$\,\Msol\ in the case of silicates and 7.5$\times 10^{-5}$\,\Msol\,\,in 
case of graphite.

\citet{Arkhipova2013} performed an analysis of the nebula spectra taken in 1990, 1992, and 2011. 
They find significant changes in the relative line intensities. The low-excitation 
$[$\ion{O}{I}$]$, $[$\ion{O}{II}$]$, and $[$\ion{N}{II}$]$ lines became stronger relative to H\,$\beta$ by 
a factor of two, while the $[$\ion{O}{III}$]$ lines weakened by a factor of $\approx 2$. Using a 
formula of \citet{Kaler1978}, they estimated that \Teff\ decreased from 1990 (\Teffw{57}) to 2011 (\Teffw{40}). 

\cite{partha1995} first discovered that the observed properties of \sao, e.g\@. the rapid 
changes in \Teff\ and the drop of luminosity, contradict with canonical post-AGB evolution. For such a 
rapid evolution, the core mass should be 0.8\,\Msol\,\,or even more \citep{partha1995, Bobrowsky1998}.
The evolutionary time scales for CSs with core masses of 0.6\,\Msol\,\,or less are predicted to be much longer \citep{bloecker1995}.

To address the evolution of the properties of SAO 244567 quantitatively for the first time, we carried out a spectral 
analysis based on all available spectra from 1988 until 2006 taken with 
IUE, 
FUSE\footnote{Far Ultraviolet Spectroscopic Explorer}, 
HST/STIS\footnote{Hubble Space Telescope / Space Telescope Imaging Spectrograph}, and 
HST/FOS\footnote{Faint Object Spectrograph}. The comparison of the results to different evolutionary models should help 
provide conclusions on the nature of \sao.

This paper is organized as follows. In \se{sect:observation}, we describe the observations. The 
spectral analysis follows in \se{sect:analysis}. 
In \se{sect:discussion}, we summarize our results and derive the distance and the mass of \sao. 
We discuss possible stellar evolutionary scenarios and compare \sao to other low-mass CSPNe. 
We conclude in \se{sect:conclusions}.

\section{Observations}
\label{sect:observation} 

\begin{figure}
  \resizebox{\hsize}{!}{\includegraphics{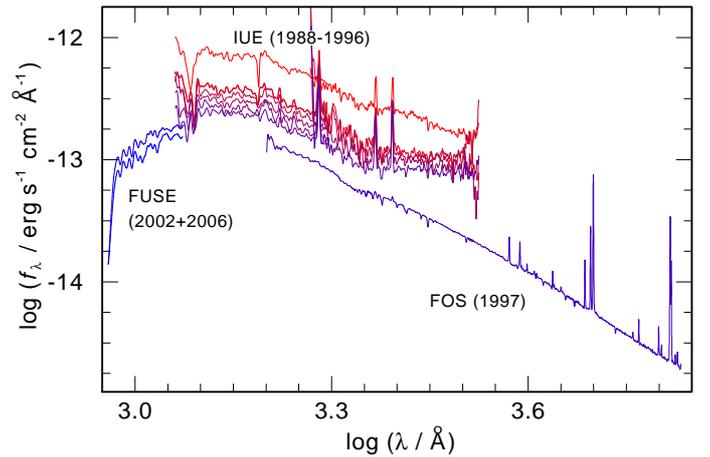}}
  \caption{All reliably flux-calibrated observations of \sao.}
  \label{fig:spectra}
\end{figure}

\sao was observed with various telescopes (Table\,\ref{tab:observations}). 
\citet{partha1995} reported the decrease in the flux level of the IUE observations by a factor of 2.83. 
We found that \sao has faded even more; e.g., we found a decrease by a factor of five by comparing the IUE observation 
from 1988 to the FOS observation in 1997. In Fig.~\ref{fig:spectra}, we show all available, reliable flux-calibrated 
observations of \sao. The STIS observations are not photometric because of the narrow slit width used and therefore 
are not shown in Fig.~\ref{fig:spectra}. Comparing the FOS observation in 1997 to the FUSE observations in 2002, we found 
that the flux must have increased slightly. Comparing the flux level of the FUSE spectrum from 2002 to the one in 
2006, we found a decrease of 15\,\%. We mention all the available spectra and their special features in \se{sect:description}. 
The determination of the interstellar reddening follows in \se{sect:ism}.

\onltab{
\onecolumn
\begin{table}
\caption{Observations of \sao.}
\label{tab:observations}
\begin{tabular}{c c c c r} 
\hline
\hline
instrument /       & ObsId & aperture/ & start time (UT) & exposure time (s) \\
telescope          &       & grating   &                  &  \\
\hline 
\noalign{\smallskip}
B \& C, ESO 1.5m  &   &   & 1990-06        & 1-600\hspace{7mm}\hbox{} \\
B \& C, ESO 1.5m  &   &   & 1992-08        & 1-600\hspace{7mm}\hbox{} \\
\hline                                                           
\noalign{\smallskip}
IUE & SWP33954LL     & large                & 1988-07-21 23:44:32 &  1800\hspace{7mm}\hbox{} \\
    & LWP13715LL     & large                & 1988-07-22 00:34:19 &  1200\hspace{7mm}\hbox{} \\
    & SWP33955LL     & large                & 1988-07-22 01:05:33 &  1800\hspace{7mm}\hbox{} \\
    & LWP22874LL     & large                & 1992-04-23 04:19:56 &   600\hspace{7mm}\hbox{} \\
    & SWP44459LL     & large                & 1992-04-23 03:52:00 &  1200\hspace{7mm}\hbox{} \\
    & LWP25397LL     & large                & 1993-04-23 05:06:09 &   600\hspace{7mm}\hbox{} \\
    & SWP47530LL     & large                & 1993-04-23 05:23:38 &  1200\hspace{7mm}\hbox{} \\
    & SWP50589LL     & large                & 1994-04-19 08:13:06 &  1200\hspace{7mm}\hbox{} \\
    & LWP27971LL     & large                & 1994-04-23 01:56:53 &   600\hspace{7mm}\hbox{} \\
    & SWP51772HL     & large                & 1994-08-11 01:52:34 & 25200\hspace{7mm}\hbox{} \\
    & SWP51852LL     & large                & 1994-08-20 09:40:09 &   900\hspace{7mm}\hbox{} \\
    & LWP30479LL     & large                & 1995-04-19 02:31:12 &   600\hspace{7mm}\hbox{} \\
    & SWP54463LL     & large                & 1995-04-19 01:59:00 &  1200\hspace{7mm}\hbox{} \\
    & SWP54465LL     & large                & 1995-04-19 08:18:31 &  1800\hspace{7mm}\hbox{} \\
    & LWP30490LL     & large                & 1995-04-20 01:45:20 &  1800\hspace{7mm}\hbox{} \\
    & SWP55690HL     & large                & 1995-08-24 18:23:29 & 22920\hspace{7mm}\hbox{} \\
    & LWP32551LL     & large                & 1996-08-10 20:08:00 &  1200\hspace{7mm}\hbox{} \\
    & LWP32552LL     & large                & 1996-08-10 21:54:07 &  1800\hspace{7mm}\hbox{} \\
    & SWP57774LL     & large                & 1996-08-10 19:31:12 &  1800\hspace{7mm}\hbox{} \\
    & SWP57775LL     & large                & 1996-08-10 20:55:02 &  3000\hspace{7mm}\hbox{} \\
\hline                                                 
\noalign{\smallskip}                                   
FOS, HST & Y3415205T    & 0$\farcs$3, G190H      & 1997-02-03 11:21:13 &   180\hspace{7mm}\hbox{} \\
         & Y341520BT    & 0$\farcs$3, G400H      & 1997-02-03 12:25:41 &   950\hspace{7mm}\hbox{} \\
         & Y341520CT    & 0$\farcs$3, G570H      & 1997-02-03 12:48:27 &  1500\hspace{7mm}\hbox{} \\
         & Y341520KT    & 0$\farcs$3, G270H      & 1997-02-03 15:38:38 &  1650\hspace{7mm}\hbox{} \\
\hline                                                           
\noalign{\smallskip}                                             
STIS, HST & O4NH01030   & 52$\arcsec$$\times$0$\farcs$05, G750M\,\,\,\,\, & 1998-03-07 22:58:48 &   708\hspace{7mm}\hbox{} \\   
     & O4NH01040   & 52$\arcsec$$\times$0$\farcs$05, G430L\,\,\,\,\,\, & 1998-03-07 23:15:55 &   840\hspace{7mm}\hbox{} \\   
     & O4NH02030   & 52$\arcsec$$\times$0$\farcs$05, G230LB\,\,      & 1999-02-05 18:13:25 &  1978\hspace{7mm}\hbox{} \\  
     & O4C562020   & 52$\arcsec$$\times$0$\farcs$05, G230L\,\,\,\,\, & 1999-04-12 01:22:42 &   900\hspace{7mm}\hbox{} \\   
     & O4C562030   & 52$\arcsec$$\times$0$\farcs$05, G140M\,\,\,\, & 1999-04-12 01:45:28 &   862\hspace{7mm}\hbox{} \\   
     & O6IH20020   & 52$\arcsec$$\times$0$\farcs$05, G430M\,\,\,\, & 2001-09-16 02:35:05 &   900\hspace{7mm}\hbox{} \\   
     & O6IH20060   & 52$\arcsec$$\times$0$\farcs$05, G430M\,\,\,\, & 2001-09-16 03:20:42 &  1080\hspace{7mm}\hbox{} \\  
     & O6IH30020   & 52$\arcsec$$\times$0$\farcs$05, G750M\,\,\,\, & 2001-09-24 08:55:01 &  1080\hspace{7mm}\hbox{} \\  
     & O6IH30060   & 52$\arcsec$$\times$0$\farcs$05, G750M\,\,\,\, & 2001-09-24 10:32:46 &  1080\hspace{7mm}\hbox{} \\  
\hline                                                           
\noalign{\smallskip}                                             
FUSE & Q3040101000 & LWRS                           & 2002-04-12 02:00:59 &  8240\hspace{7mm}\hbox{} \\
     & Q3040102000 & LWRS                           & 2002-06-16 22:12:47 & 15268\hspace{7mm}\hbox{} \\  
     & U1093201000 & LWRS                           & 2006-06-23 04:02:36 & 15824\hspace{7mm}\hbox{} \\  
     & U1093202000 & LWRS                           & 2006-06-24 11:12:43 &  2862\hspace{7mm}\hbox{} \\
\hline                                                        
\end{tabular}
\end{table}
\twocolumn
}

\subsection{Description of the spectra}
\label{sect:description}

IUE spectra were taken from 1988 to 1996. All these observations but 
SWP51772 and SWP55690 were obtained in low-resolution mode. The main features in the spectra are 
nebular emission lines (from 1992 on) and the \ion{C}{IV} and \ion{N}{V} resonance doublets that are blue 
shifted and show P-Cygni profiles (\se{sect:wind}).

The HST/FOS spectra were obtained in 1997 using the G190H, G270H, G400H, and G570H gratings 
(total exposure time $\approx$ 4600\,s, ProgID 6039). The observations cover a wavelength 
range from $\approx$ 1600 to 6800\,\AA.

\sao was observed with HST/STIS in 1998 (ProgID 7652), 1999 (ProgID 7652, 7653), and 
2001 (ProgID 8929) using the 52\arcsec $\times$ 0$\farcs$05 aperture. These observations cover a total wavelength 
range from 1540 to 6000\,\AA\,\,and from 9050 to 9650\,\AA. 
In the optical wavelength range (3200 to 6800\,\AA), the HST/FOS and HST/STIS spectra 
show many nebular emission lines, but also some photospheric \ion{He}{II} lines (partially blended by the nebula 
emission lines). The NUV spectra turned out to be more useful for the spectral analysis, because of the 
\ion{He}{II} lines ($n\,\rightarrow\,n'=3\,\rightarrow\,6, ..., 14$), which could be used to determine 
\logg, \Teff, and the H/He ratio. We identified \ion{C}{III} and \ion{C}{IV} lines which allowed us a more precise 
determination of \Teff by evaluation of the \ion{C}{III}\,/\,\ion{C}{IV} ionization equilibrium 
(see Sect. \ref{sect:tefflogg}). Apart from the 
photospheric lines, we found some interstellar \ion{Fe}{II}, \ion{Mg}{II}, and \ion{Mn}{II} lines.

FUV spectra were taken with FUSE in 2002 and 2006 (total exposure time: 
31\,ksec, ProgIDs Q304 and U109) using the LWRS aperture. These spectra show many interstellar lines, 
but also photospheric \ion{H}{I}, \ion{He}{II}, \ion{C}{III}, \ion{C}{IV}, \ion{N}{III} (2002 only), 
\ion{N}{IV}, \ion{O}{III}, \ion{O}{IV}, \ion{Si}{IV}, \ion{P}{V}, \ion{S}{IV}, \ion{S}{V}, \ion{S}{VI}, 
\ion{Fe}{V}, and \ion{Fe}{VI} lines. Furthermore, the \ion{O}{VI} doublet lines show weak P-Cygni profiles. 
We checked the FUSE observations for relative motion of interstellar and photospheric lines, but we could 
not find any hint of velocity shifts either by comparing the observation from 2002 and 2006, or in the 
individual exposures of the corresponding years. This indicates (except in case of a very high inclination angle) 
the probability of a close companion is rather unlikely. 
All these space-based observations were retrieved from the MAST archive.

\subsection{Interstellar reddening}
\label{sect:ism}

\begin{figure}
  \resizebox{\hsize}{!}{\includegraphics{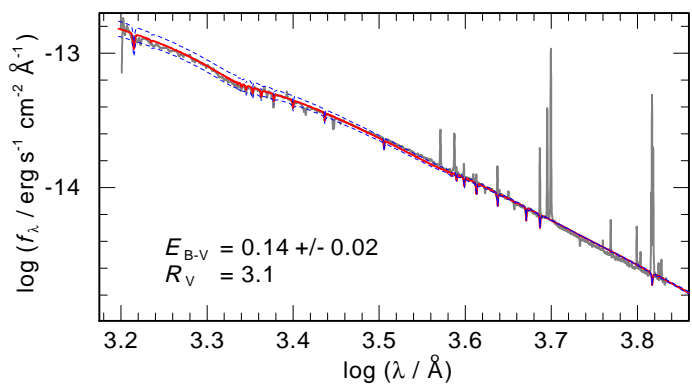}}
  \caption{Example for the determination of \ebv for \sao. The FOS observation (gray) is compared to TMAW model fluxes 
    with different corrections for the interstellar reddening. The thick (red in online version) line represents the TMAW model 
    corrected with $E_{B-V}=0.14$, the dashed lines (blue) indicate the error limits.} 
  \label{fig:EBV}
\end{figure}

We derived the interstellar reddening for each observation. 
The flux shape of the IUE, STIS, and FOS spectra can only be reproduced using the LMC reddening 
law of \citep{howarth1983}. The FUSE observations could be fitted with the reddening law of \citet{fitzpatrick1999}. 
Figure~\ref{fig:EBV} shows the determination of $E_{B-V}$ for the FOS observation. Since the impact of the interstellar 
reddening is negligible in the infrared, the TMAW model flux of our best fit model to the FOS observation (\Teffw{50}, 
\loggw{5.5}, \se{sect:tefflogg}) was firstly normalized to the reddest part of the spectrum. Then, we corrected the model 
flux for different values of $E_{B-V}$ to fit the model flux to the observation.

We confirm the variations in $E_{B-V}$ found by \citet{Arkhipova2013}, who measured the reddening from the 
H\,$\beta$ line intensity, using the simple relation, $E_{B-V}=0.68 \times c(H\,\beta)$, with the 
extinction coefficient c(H\,$\beta$). In Table\,\ref{tab:ebv}, we compare our values with those of other authors.

\begin{table}
	\centering
	\caption{Interstellar reddening of \sao. Typical errors for $E_{B-V}$ are $\pm0.02$.}
        \label{tab:ebv}
        \begin{tabular}{c c l l}
	\hline
	\hline
         Year  & Telescope  & $E_{B-V}$  &  Method  \\
        \hline
        \noalign{\smallskip}
        1980   &  61cm ESO  &  0.18   & UBV photometry$^{\mathrm{d}}$  \\
        1988   &  IUE       &  0.18   & LMC reddening law$^{\mathrm{a}}$  \\
        1988   &            &  0.20   & Seaton reddening law$^{\mathrm{c}}$  \\
        1990   &  1.5m ESO  &  0.18   & H\,$\beta$ line intensity$^{\mathrm{b}}$  \\  
               &            &  0.18   & H\,$\beta$ line intensity$^{\mathrm{c}}$  \\  
               &            &  0.14   & Balmer decrement$^{\mathrm{3}}$  \\  
        1992   &  1.5m ESO  &  0.13   & H\,$\beta$ line intensity${\mathrm{^b}}$  \\  
        1993   &  IUE       &  0.17   & LMC reddening law$^{\mathrm{a}}$  \\  
        1994   &  IUE       &  0.19   & LMC reddening law$^{\mathrm{a}}$  \\  
        1995   &  IUE       &  0.20   & LMC reddening law$^{\mathrm{a}}$  \\  
        1996   &  IUE       &  0.20   & LMC reddening law$^{\mathrm{a}}$  \\  
        1997   &  FOS       &  0.14   & LMC reddening law$^{\mathrm{a}}$  \\  
        1999   &  STIS      &  0.18   & LMC reddening law$^{\mathrm{a}}$  \\  
        2002   &  FUSE      &  0.11   & Fitzpatrick reddening law$^{\mathrm{a}}$  \\  
        2006   &  FUSE      &  0.11   & Fitzpatrick reddening law$^{\mathrm{a}}$  \\  
        2011   &  1.9m SAAO &  0.24   & H\,$\beta$ line intensity$^{\mathrm{b}}$ \\  
   	\hline 
	\end{tabular}
\tablefoot{~\\
\tablefoottext{a}{this work}
\tablefoottext{b}{\citet{Arkhipova2013}}
\tablefoottext{c}{\citet{partha1993}}
\tablefoottext{d}{\citet{kozok1985b}}
} 
\end{table}

\section{Spectral analysis}
\label{sect:analysis}

The procedure of our spectral analysis is the following. First, we derived \Teff, 
\logg, and the abundances (or at least upper limits) of the elements H, He, C, N, O, Ne, 
Si, P, S, Fe, and Ni with the T{\"u}bingen NLTE model-atmosphere package 
\citep[TMAP\footnote{\url{http://astro.uni-tuebingen.de/~TMAP}},][]{werneretal2003,rauchdeetjen2003} 
from the FOS, STIS, and FUSE spectra. The mass-loss rate of \sao during those years, in which these 
observations were taken, is negligible and therefore it is justified to use a code for hydrostatic atmospheres. 
We compared the PoWR\footnote{\url{http://www.astro.physik.uni-potsdam.de/~wrh/PoWR/}} 
\citep[Potsdam Wolf-Rayet model-atmosphere code,][]{graefeneretal2002, hamannetal2003, hamannetal2004} 
and TMAP models for the FUSE observations and found that apart from the O\,{\sc vi} P Cygni profiles, the spectral 
lines are reproduced equally well. The TMAP analysis is described in \se{sect:model}. To study the mass-loss rates 
and terminal wind velocities from the P Cygni profiles found in the IUE and FUSE spectra and to determine \Teff\ in 
the years of the IUE observations, we used PoWR. The description of this analysis can be found in \se{sect:wind}.

\subsection{TMAP model atmospheres}
\label{sect:model}

We used TMAP to compute non-LTE, plane-parallel, fully metal-line blanketed model atmospheres 
in radiative and hydrostatic equilibrium. The final models included opacities of the elements 
H, He, C, N, O, Ne, Si, P, S, Fe, and Ni. The model atoms for this analysis were taken
from the T{\"u}bingen model-atom database TMAD\footnote{\url{http://astro.uni-tuebingen.de/~TMAD}} 
and calculated (Fe, Ni) via the T{\"u}bingen iron-group opacity interface 
TIRO\footnote{\url{http://astro.uni-tuebingen.de/~TIRO}} \citep{ringatPhD2013}.
It has been developed recently in the framework
of the Virtual Observatory (\emph{VO}\footnote{\url{http://www.ivoa.net}})
and is provided as a registered service by the \emph{German Astrophysical Virtual Observatory}
(\emph{GAVO}\footnote{\url{http://www.g-vo.org}}).
The statistics of the model atoms used in this analysis are summarized in 
Table\,\ref{tab:statistics}. All spectral energy distributions that were calculated for this analysis 
are available via the registered GAVO service TheoSSA\footnote{\url{http://dc.g-vo.org/theossa}}. 

\onltab{
\begin{table}
        \centering
        \caption{Statistics of our model atoms. For iron and nickel, we used a statistical superline/superlevel approach \citep{rauchdeetjen2003} in the subsequent NLTE level population calculations. The original number of the so-called ``sample lines'' from Kurucz's line list, which are combined to super-lines, is also given.}
  \label{tab:statistics}
\begin{tabular}{r l c c r}
        \hline
        \hline
  \multicolumn{2}{c}{ion}  &  NLTE levels &  lines  & sample lines \\ 
\hline
\noalign{\smallskip}
H  & {\sc i}   &      5 &     10 &           \\
   & {\sc ii}  &      1 &     $-$&           \\
\noalign{\smallskip}
He & {\sc i}   &     29 &     69 &           \\
   & {\sc ii}  &     20 &    190 &           \\
   & {\sc iii} &      1 &     $-$&           \\
\noalign{\smallskip}
C  & {\sc iii} &     58 &    329 &           \\
   & {\sc iv}  &     54 &    295 &           \\
   & {\sc v}   &      1 &     0  &           \\
\noalign{\smallskip}
N  & {\sc iii} &     34 &    129 &           \\
   & {\sc iv}  &     90 &    546 &           \\
   & {\sc v}   &     54 &    297 &           \\
   & {\sc vi}  &      1 &     0  &           \\
\noalign{\smallskip}
O  & {\sc iii} &     72 &    322 &           \\
   & {\sc iv}  &     38 &    173 &           \\
   & {\sc v}   &     53 &    255 &           \\
   & {\sc vi}  &     54 &    291 &           \\
   & {\sc vii} &      1 &      0 &           \\
\noalign{\smallskip}
Ne & {\sc ii}  &     14 &     19 &           \\
   & {\sc iii} &     14 &     17 &           \\
   & {\sc iv}  &     14 &     24 &           \\
   & {\sc v}   &     14 &     18 &           \\
   & {\sc vi}  &      1 &      0 &           \\
\noalign{\smallskip}
Si & {\sc iii} &     17 &     28 &           \\
   & {\sc iv}  &     16 &     44 &           \\
   & {\sc v}   &     25 &     59 &           \\
   & {\sc vi}  &      1 &      0 &           \\
\noalign{\smallskip}
P  & {\sc iv}  &     15 &      9 &           \\
   & {\sc v}   &     18 &     12 &           \\
   & {\sc vi}  &      1 &      0 &           \\
\noalign{\smallskip}
S  & {\sc iv}  &     20 &     25 &           \\
   & {\sc v}   &     23 &     33 &           \\
   & {\sc vi}  &     30 &     42 &           \\
   & {\sc vii} &      1 &      0 &           \\
\noalign{\smallskip}
Fe & {\sc iv}  &      5 &    12  &  3\,102\,371  \\
   & {\sc v}   &      7 &    25  &  3\,266\,247  \\
   & {\sc vi}  &      7 &    25  &   991\,935  \\
   & {\sc vii} &      1 &     0  &           \\
\noalign{\smallskip}
Ni & {\sc iv}  &      5 &    14  &  2\,512\,561  \\
   & {\sc v}   &      7 &    25  &  2\,766\,664  \\
   & {\sc vi}  &      7 &    22  &  7\,408\,657  \\
   & {\sc vii} &      1 &     0  &           \\
\hline           
     & \multicolumn{1}{r}{total} &    467  &  1022 & 20048435  \\
\hline
\end{tabular}
\end{table}
}

In a standard procedure \citep[e.g\@.][]{ziegler2012}, we modeled both, the photospheric and the interstellar 
line-absorption spectrum, to correctly identify the stellar lines. We employed the OWENS program to model the 
ISM line absorption with several ISM clouds with different parameters (radial and turbulent velocities, temperatures, 
and column densities of the individual ions). In the following, we describe our photospheric analysis.

\begin{figure*}[ht]
 \centering
  \includegraphics[width=\textwidth]{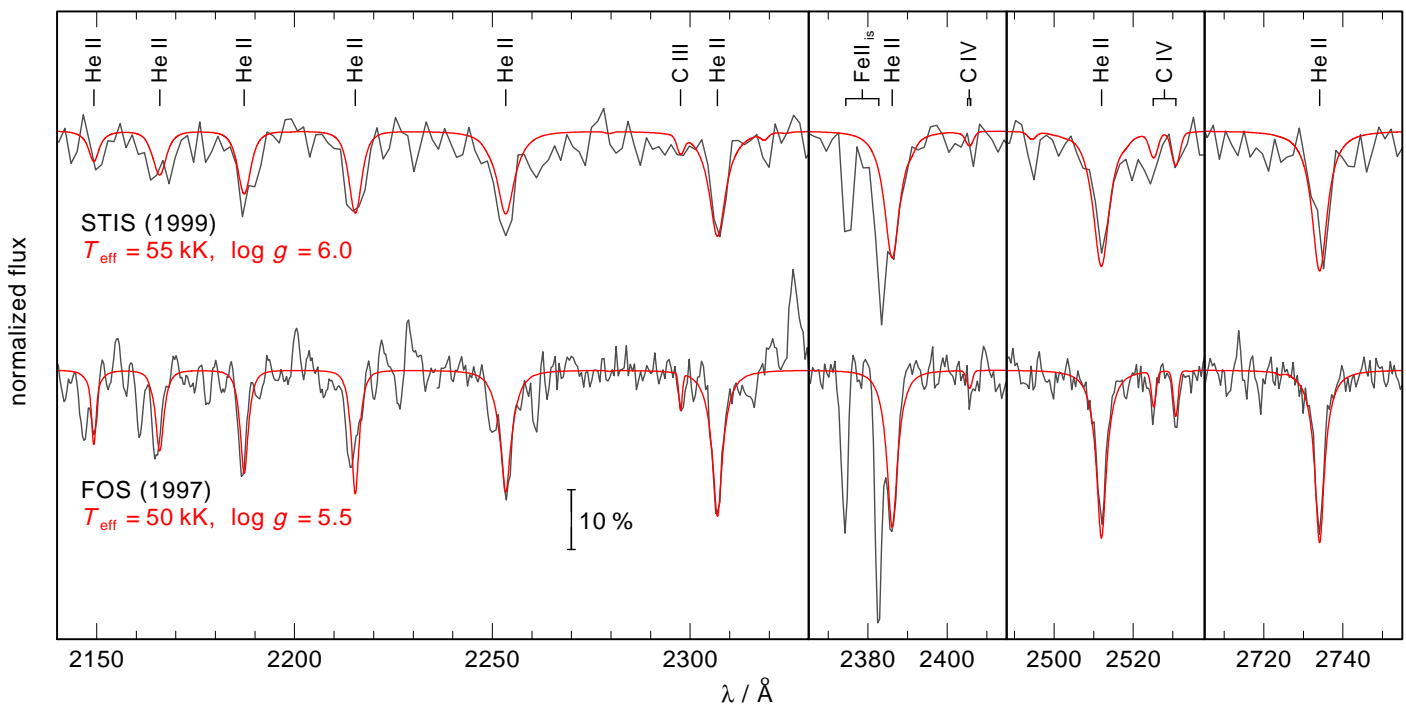}
  \caption{Comparison of our best-fit TMAP models (black, red in online version) to the FOS and STIS observation (gray). He\,{\sc ii}, C\,{\sc iii} - {\sc iv}, and interstellar Fe\,{\sc ii} lines are marked. The vertical bar indicates 10\,\% of the continuum flux.}
  \label{fig:fosstis}
\end{figure*}

\subsection{Effective temperature, surface gravity, H/He ratio, C abundance}
\label{sect:tefflogg}

Based on the IUE spectrum in 1995, \citet{partha1995} estimated \Teff\,$\approx$\,55\,kK. 
Starting with this value, we computed a grid of H+He composed model atmospheres 
(\Teffw{40  - 70}, 10\,kK steps),
 \loggw{4.0 - 7.0}, 0.5 steps), 
H/He abundance ratios (by mass) of 9/1, 3/1 (i.e\@. solar), and 2/3 
for a preliminary \Teff and \logg determination.
The synthetic spectra of these models were then compared to the observed He line
profiles in the STIS and FOS spectra.

\Ionw{He}{2}{2723.2} is much too strong in all H/He\,$=$\,2/3 models so 
we can rule out He enrichment. The H/He\,$=$\,9/1 models show that \sao is not
He-poor because it is not possible to match the strengths and wings of the \ion{He}{II}
lines. All line profiles are reproduced well with a solar H/He ratio
at \loggw{6.0} and \Teffw{50, 60, 70} for the STIS spectrum. For the FOS 
observations the best fit was found for \loggw{5.5} and \Teffw{40, 50, 60}.
We adopted this abundance ratio for our analysis. 

We identified several \ion{C}{III} and \ion{C}{IV} lines in the FUSE, STIS and FOS spectra.
Consequently, we included C in our models to evaluate the 
\ion{C}{III}\,/\,\ion{C}{IV} ionization equilibrium and to determine the C abundance. 
For the determination of the C abundance, 
we used
\Ionww{C}{3}{1165.6-1165.9, 1174.9-1176.4} and 
\Ionww{C}{4}{1107.6-1108.0, 1168.8-1169.0} 
in the FUSE spectrum and 
\Ionw{C}{3}{2297.6} and
\ion{C}{IV} $\lambda\lambda$ $2336.7-2337.1$, $2405.1-2405.9$, $2524.41-2530.7$\,\AA\ 
in the STIS and FOS spectra. The \Ionww{C}{4}{1548.2, 1550.8} resonance doublet is contaminated by the 
respective interstellar lines, hence not suited to determine the C abundance. The best agreement with 
the observation was found at  $\mathrm{C} = 5.0\times 10^{-4}$, which is 0.2\,$\times$ solar (solar values 
according to \citealt{Asplund2009}) in all spectra. Simultaneously, we refined our model grid (\Teff step 5\,kK). 
In Fig.~\ref{fig:fosstis}, we show the best-fit models for the FOS observation in 1997 and the STIS observation in 1999. We discovered that the ionization equilibrium and also the surface gravity have changed with time. While we determined \Teffw{50} and \loggw{5.5} for the FOS observation, the STIS observation is best reproduced by \Teffw{55} and \loggw{6.0}.

In the FUSE spectra, we evaluated ionization equilibria of 
\ion{C}{III}\,/\,\ion{C}{IV}, 
\ion{N}{III}\,/\,\ion{N}{IV} (FUSE 2002 only), 
\ion{O}{III}\,/\,\ion{O}{IV}, and
\ion{S}{IV}\,/\,\ion{S}{V}\,/\,\ion{S}{VI} to derive \Teff. 
In Fig.~\ref{fig:Teff}, we show the different ionization equilibria of the FUSE observation in 2002. The model with \Teffw{60} matches the observations. In the 55\,kK model, the \ion{O}{III} and \ion{S}{IV} lines are too strong, whereas the \ion{S}{VI} line is too weak. In the 65\,kK model, the \ion{N}{III}, \ion{O}{III}, \ion{S}{IV}, and \ion{C}{III} lines are too weak and the \ion{S}{VI} line is too strong. Comparing the FUSE observations from 2002 and 2006, we found that the \ion{O}{III} and \ion{S}{IV} lines became slightly stronger. Figure~\ref{fig:Teff2006} shows that the best agreement in 2006 is obtained for \Teffw{55}.
The value of \loggw{6.0\pm0.5} was confirmed using the \ion{He}{II} line wings. In Table\,\ref{tab:wind}, we list the values of \Teff\ and \logg\ that we found from this analysis.

\onlfig{
\begin{landscape}
\addtolength{\textwidth}{6.3cm} 
\addtolength{\evensidemargin}{-3cm}
\addtolength{\oddsidemargin}{-3cm}
\begin{figure*}[ht]
 \centering
  \includegraphics[trim=0 0 0 -120,height=15cm,angle=0]{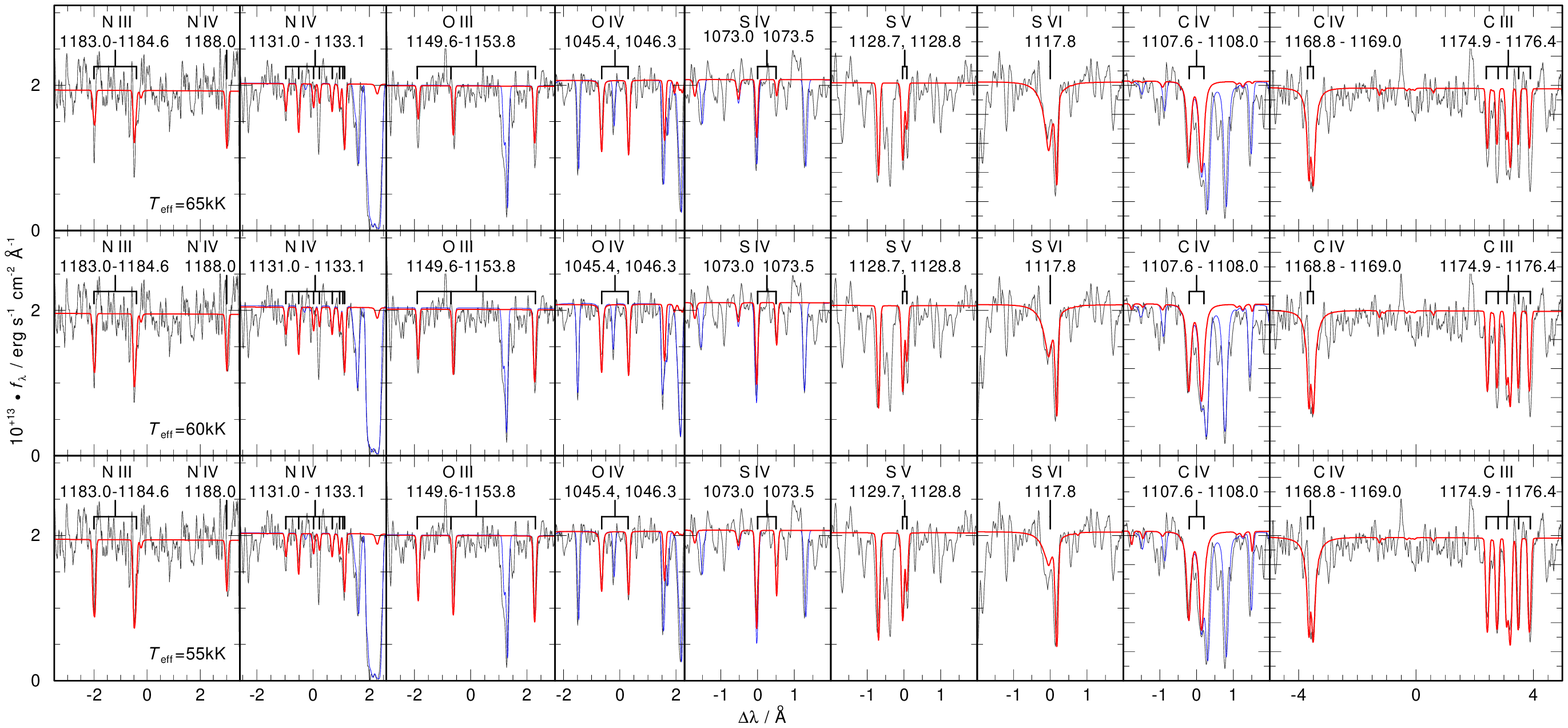}
  \caption{Determination of \Teff using the ionization equilibria of N, O, S and C in the FUSE spectrum taken in 2002. The red (thick) lines indicate the pure stellar spectrum, the blue (thin) lines also include the interstellar lines.           
           N\,{\sc iii} - {\sc iv}, O\,{\sc iii} - {\sc v}, S\,{\sc iv} - {\sc vi} 
           and C\,{\sc iii} - {\sc iv} lines are marked.
            }
  \label{fig:Teff}
\end{figure*}
\end{landscape}
}

\onlfig{
\begin{landscape}
\addtolength{\textwidth}{6.3cm} 
\addtolength{\evensidemargin}{-3cm}
\addtolength{\oddsidemargin}{-3cm}
\begin{figure*}[ht]
 \centering
  \includegraphics[trim=0 0 0 -120,height=15cm,angle=0]{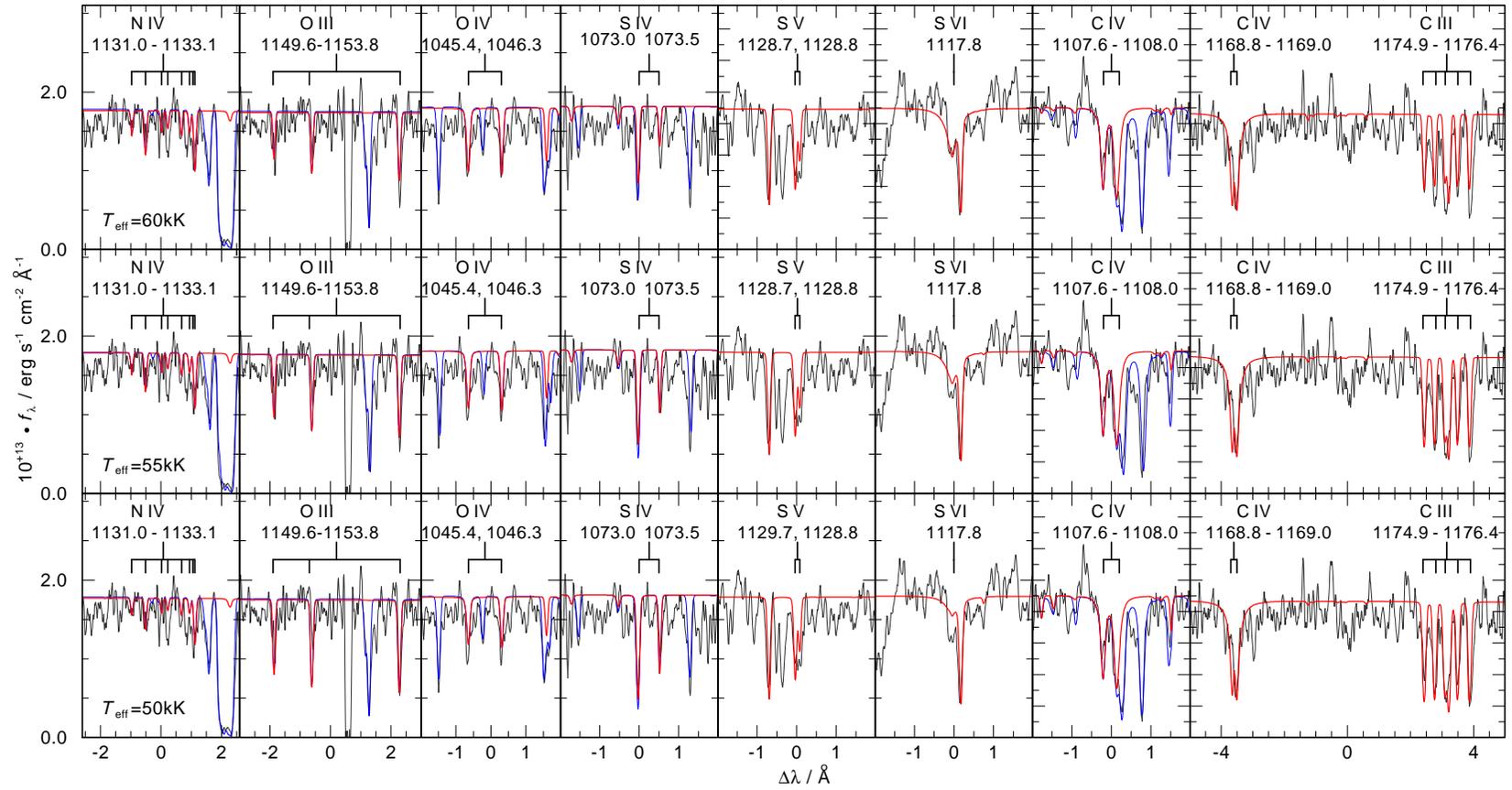}
  \caption{Like Fig.~\ref{fig:Teff}, but for the year 2006.}
  \label{fig:Teff2006}
\end{figure*}
\end{landscape}
}

\begin{table}
	\centering
	\caption{Temporal evolution of the atmospheric parameters as derived from our spectral analysis. Typical errors are $\Delta$\Teff $=\pm$ 5\,kK, $\Delta$\logg $=\pm$ 0.5, $\Delta$\Mdot $=\pm$ 0.2\,dex, and $\Delta v_\infty$ $=\pm$200\,km/s.} 

        \label{tab:wind}
        \begin{tabular}{c c c r@{.} l c c}
	\hline
	\hline
        \noalign{\smallskip}
        Year   & \Teff\ & $\log g$    & \multicolumn{2}{l}{$\log$ \Mdot}   & $v_\infty$   & Code \\  
               &  [kK]  & [cm\,/\,s$^2$] & \multicolumn{2}{l}{[\Msol/yr]}     & [km/s]       &  \\  
         \hline
        1988   &  38 & 4.8 & $-$9&0 & 1800  & \small{PoWR} \\
        1992   &  43 & 5.0 & $-$9&0 & 1800  & \small{PoWR} \\  
        1993   &  44 & 5.0 & $-$9&1 & 2100  & \small{PoWR} \\  
        1994   &  48 & 5.2 &$-$10&0 & 2100  & \small{PoWR} \\  
        1995   &  50 & 5.2 &$-$10&0 & 2400  & \small{PoWR} \\  
        1996   &  50 & 5.2 &$-$10&0 & 2800  & \small{PoWR} \\  
        1997   &  50 & 5.5 &  \multicolumn{2}{l}{}    &   & \small{TMAP}   \\  
        1999   &  55 & 6.0 &  \multicolumn{2}{l}{}    &   & \small{TMAP}   \\  
        2002   &  60 & 6.0 &$-$11&3 & 2800  & \small{TMAP/PoWR} \\  
        2006   &  55 & 6.0 &$-$11&6 & 2800  & \small{TMAP/PoWR} \\  
   	\hline 
	\end{tabular}
\end{table}

\subsection{Element abundances}
\label{sect:abund}
For determining the abundances of N, O, Si, P, S, and Ni, we used the respective lines found 
in the FUSE spectra. To derive upper limits, we compared our models to FOS spectra. 
For the determination of the element abundances,
we adopted the values of \Teff and \logg found for each epoch (\se{sect:tefflogg}). 
The He and C abundances were already achieved as a byproduct
of the \Teff and \logg determination (see above).
To calculate the \sao model grids in a reasonable time,
we only included H, He, C, N, and O and added
the trace elements Ne, Mg, Si, P, S, Fe, and Ni in a subsequent
line-formation calculation; i.e\@., we kept the atmospheric structure fixed
and calculated NLTE occupation numbers only for newly introduced species. 
This is justified by the comparison of temperature structures of a
HHeCNO and a complete, final HHeCNONeSiPSFeNi model, with \Teffw{60} and \loggw{6.0}, where the 
atmospheric structure is calculated with all these elements. Figure ~\ref{fig:Tstruc} demonstrates
that the temperature structures are almost identical in the line-formation region 
($-4\,\, \sla \,\,\log m \,\,\sla \,\,1$). All abundances found from line-formation calculations 
are verified with our final models.

\onlfig{
\begin{figure}
  \resizebox{\hsize}{!}{\includegraphics{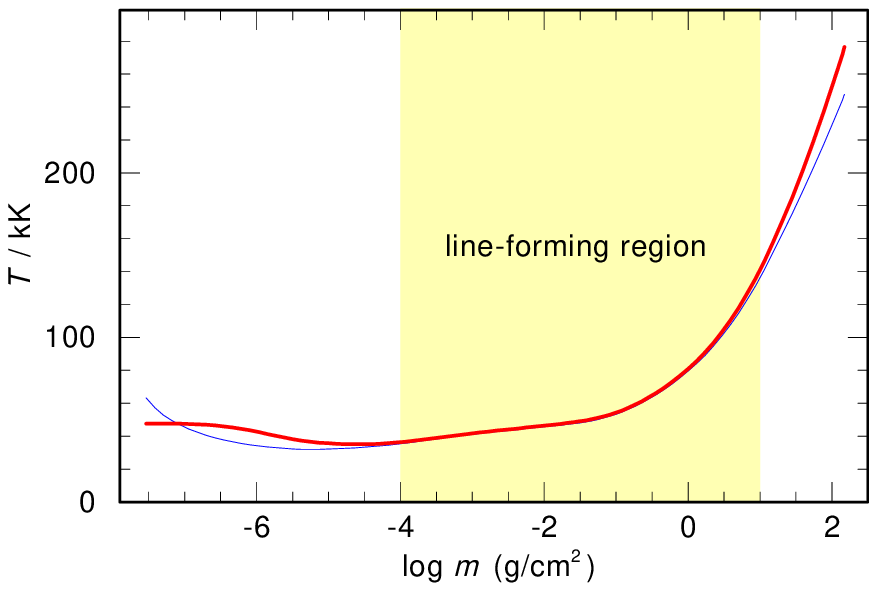}}
  \caption{Comparison of the temperature structures of a
           HHeCNO (blue, thin) and the final HHeCNONeSiPSFeNi model (red, thick) with \Teffw{60} and \loggw{6.0}.} 
  \label{fig:Tstruc}
\end{figure}
}

\paragraph{Nitrogen and oxygen}
\label{sect:cno}
The N abundance of \sao is solar ($\mathrm{N} = 6.9\times 10^{-4}$). 
It was measured using 
\Ionww{N}{3}{1183.0-1184.6}, 
\Ionww{N}{4}{1123.5-1133.1}, and \Ionw{N}{4}{1183.0} 
lines in the FUSE spectrum.
The O abundance was derived from the 
\Ionww{O}{3}{1149.6-1153.8}, and 
\ion{O}{IV} $\lambda\lambda$ $1045.4-1050.5$, $1167.8-1168.0$\,\AA\ 
lines identified in the FUSE spectrum (Fig.~\ref{fig:Teff}). 
We found a solar value ($\mathrm{O} = 5.6\times 10^{-3}$).

\paragraph{Neon, sodium, magnesium, silicon, phosphor, and sulfur}
\label{sect:nenamgsip}
The only Ne line that was prominent in our models is
\Ionw{Ne}{3}{2678.7}. It would emerge from the
STIS and FOS spectra above an upper abundance limit of five times the solar value 
($\mathrm{Ne}\,\sla\,6.2\times 10^{-3}$).
Using \Ionww{Na}{3}{2012.5, 2031.8}, we derive an upper limit for sodium of ten times the solar value 
($\mathrm{Na}\,\sla\,2.9\times 10^{-4}$).
For magnesium, we derive an upper limit (solar, $\mathrm{Mg}\,\sla\,7.1\times 10^{-4}$) using
\Ionww{Mg}{3}{2134.7, 2178.4}.
For the determining the Si abundance we used the 
\Ionw{Si}{3}{1113.2} 
and
\Ionw{Si}{4}{1128.3} lines. 
All other Si lines are contaminated by interstellar absorption lines. 
As illustrated in Fig.~\ref{fig:Si}, the best fit is obtained at a solar Si abundance ($\mathrm{Si} = 6.7\times 10^{-4}$). 
For lower/higher Si abundances the modeled Si lines are too weak/strong.
\Ionww{P}{5}{1118.0, 1128.0} were used to determine the P abundance (Fig.~\ref{fig:P}) 
which is 0.1\,$\times$ solar ($\mathrm{P} = 5.8\times 10^{-7}$). 
We identified 
\Ionww{S}{4}{1073.0, 1073.5}, 
\Ionw{S}{5}{1039.9}, and
\Ionw{S}{6}{1117.8} and determined the S abundance to be 0.5$\times$ solar 
($\mathrm{S} = 1.6\times 10^{-4}$).

\onlfig{
\begin{figure}
  \resizebox{\hsize}{!}{\includegraphics{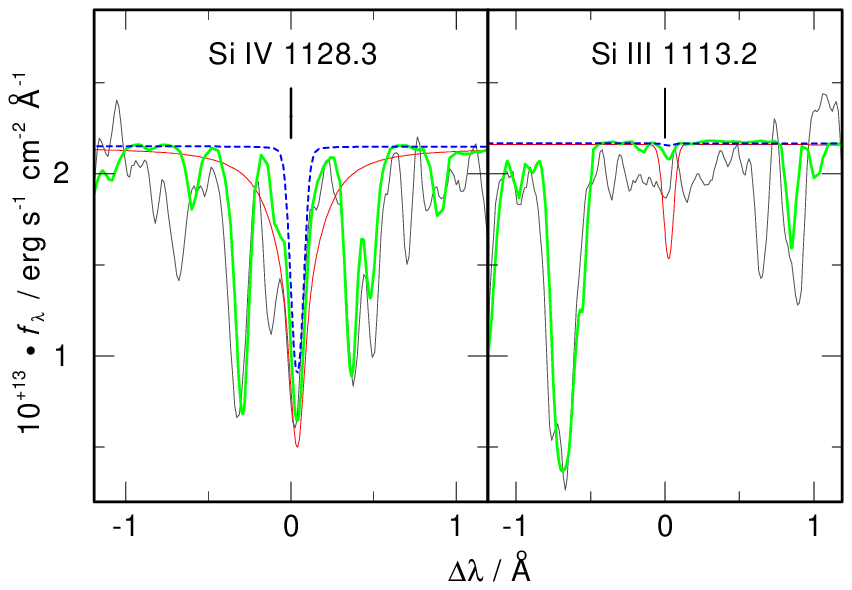}}
  \caption{Determination of the Si abundance. Synthetic line profiles calculated from 
           models with 0.1\,$\times$ solar (blue, dashed), 
                                     solar (green, light gray, including ISM absorption lines), and 
                        10\,$\times$ solar (red, dark) Si abundance
           are compared with the observed \ion{Si}{IV} (left) and
                                          \ion{Si}{III} line profiles.}
   \label{fig:Si}
\end{figure}
}

\onlfig{
\begin{figure}
  \resizebox{\hsize}{!}{\includegraphics{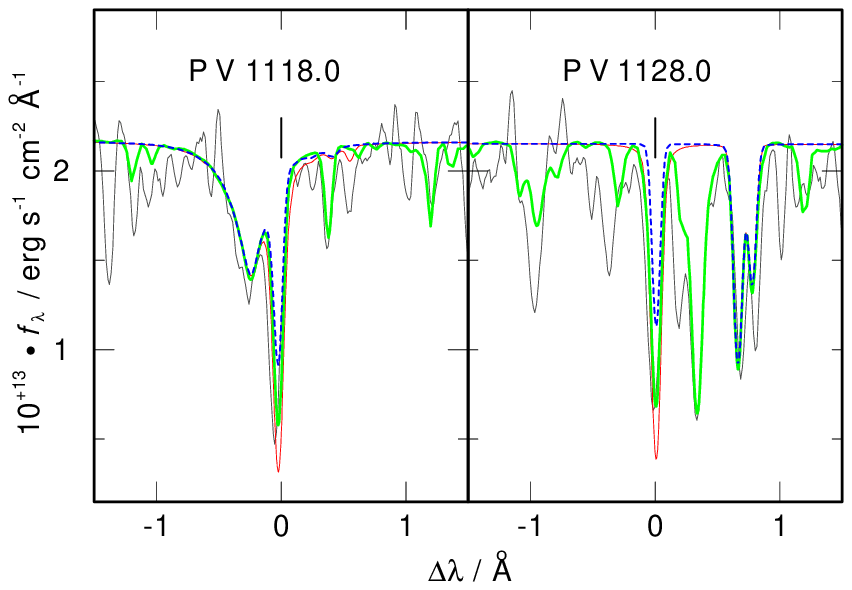}}
  \caption{Determination of the P abundance. Synthetic line profiles calculated from
           models with 0.01$\times$ solar (blue, dashed), 
                        0.1$\times$ solar (green, light gray, including ISM absorption lines), and 
                                    solar (red, dark) P abundance
           are compared with the observed \ion{P}{V} line profiles.} 
  \label{fig:P}
\end{figure}
}

\paragraph{Iron and nickel}
\label{sect:feni}

Some Fe lines could be identified in the FUSE spectra (\Ionww{Fe}{5}{999.6-1001.1}, 
\Ionww{Fe}{6}{1165.7, 1167.7}). The Fe abundance was found to be solar 
($\mathrm{Fe}\,\sla\,1.3\times 10^{-3}$). The \Ion{Fe}{5}/\Ion{Fe}{6} ionization equilibrium 
also confirms \Teffw{60} (Fig.~\ref{fig:FeNi}). The quality of the available FUSE spectra is 
not sufficient to unambiguously identify individual Ni lines. Therefore, we could only derive a 
(solar) upper limit for the Ni ($\mathrm{Ni}\,\sla\,7.1\times 10^{-5}$) abundance (Fig.~\ref{fig:FeNi}). 
We summarize the abundances in Table\,\ref{tab:Abund}.
Their error limits are $\pm 0.3$\,dex.

\onlfig{
\begin{figure*}[ht!]
 \centering
  \includegraphics[width=18cm]{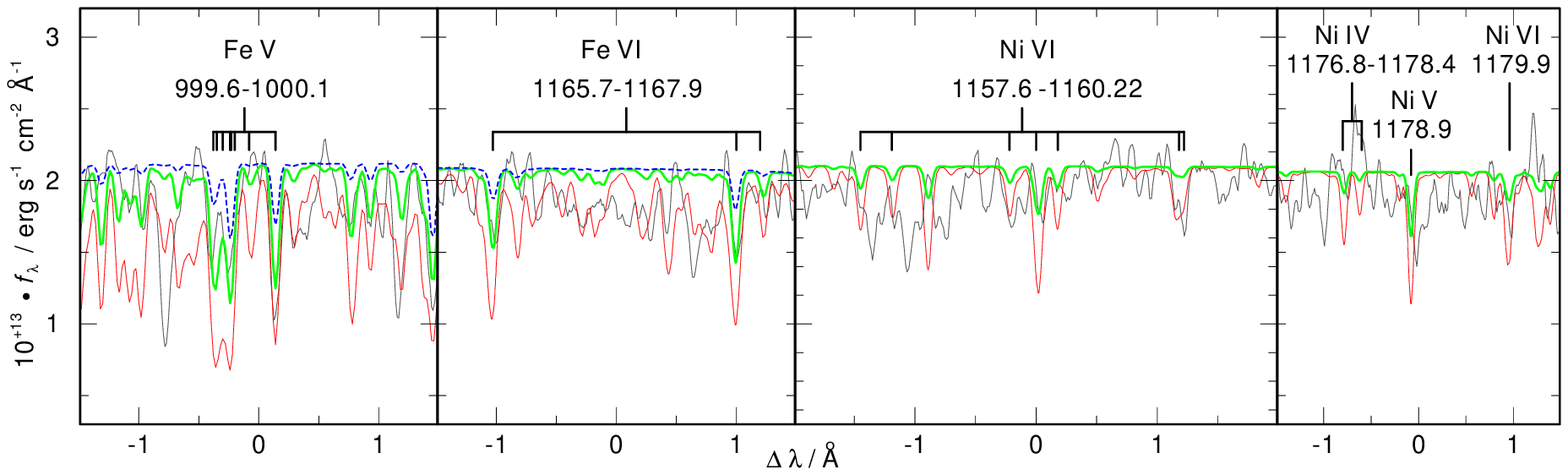}
  \caption{Determination of the Fe abundance and the upper limit for the Ni abundance. Synthetic spectra calculated from
           models with 0.1\,$\times$ solar (blue, dashed), solar (green, light) and 
           10\,$\times$ solar (red, dark) Fe and Ni abundances are compared with the observation.}
  \label{fig:FeNi}
\end{figure*}
}

\begin{table}
	\centering
	\caption{Photospheric abundances of \sao. 
                 [X] denotes log (abundance / solar abundance). $\log \varepsilon_\mathrm{X}$ are normalized to $\log \sum\mu_\mathrm{X}\varepsilon_\mathrm{X} = 12.15$, where $\mu_\mathrm{X}$ is the atomic weight.}
  \label{tab:Abund}
        \begin{tabular}{l r@{.} l r@{.} l r@{.} l r@{.} l}
	\hline
	\hline
                  & \multicolumn{2}{l}{$\log$ mass}  &  \multicolumn{2}{l}{$\log$ number} &\multicolumn{2}{l}{ }               &\multicolumn{2}{l}{ }         \\
         Element  & \multicolumn{2}{l}{fraction}     &  \multicolumn{2}{l}{fraction}      &\multicolumn{2}{l}{$\log\epsilon$}  &  \multicolumn{2}{c}{[X]} \\ 
        \hline
        \noalign{\smallskip}
                H           &  $-$0&13 &  $-$0&04  & 12&01      &     0&00   \\
                He          &  $-$0&60 &  $-$1&11  & 10&94      &     0&00   \\
                C           &  $-$3&30 &  $-$4&28  &  7&77      &  $-$0&68   \\
                N           &  $-$3&16 &  $-$4&21  &  7&84      &     0&00   \\
                O           &  $-$2&25 &  $-$3&35  &  8&70      &     0&00   \\
                Ne          & $<-$2&21 &  $<-$3&41 &  $<$\,8&64 &  $<$\,0&70 \\
                Na          & $<-$3&53 &  $<-$4&80 &  $<$\,7&26 &  $<$\,1&00 \\
                Mg          & $<-$3&15 &  $<-$4&44 &  $<$\,7&61 &  $<$\,0&00 \\
                Si          &  $-$3&18 &  $-$4&53  &  7&52      &     0&00   \\
                P           &  $-$6&24 &  $-$7&63  &  4&42      &  $-$1&00   \\
                S           &  $-$3&81 &  $-$5&22  &  6&83      &  $-$0&30   \\
                Fe          &  $-$2&89 &  $-$4&54  &  7&51      &     0&00   \\
                Ni          & $<-$4&15 &  $<-$5&82 &  $<$\,6&23 &  $<$\,0&00   \\   	
   	\hline 
	\end{tabular}
\end{table}

\subsection{Stellar wind}
\label{sect:wind}

\citet{partha1993} found that in the IUE spectra of 1988 and 1992 the 
\ion{C}{IV} and \ion{N}{V} resonance doublets are blue-shifted and they show P-Cygni profiles. 
Looking more carefully, we found that all the spectra show P-Cygni profiles in 
these lines. While in 1988 the P-Cygni profile of \ion{N}{V} is quite weak and \ion{C}{IV} strong, 
this changes from 1992 on, which indicates that the star's \Teff has increased. 
Furthermore, we found in the FUSE spectra that the \Ionww{O}{6}{1031.9, 1037.6} resonance 
lines show P-Cygni profiles, too.

To measure the mass-loss rate and \Teff of \sao, we used PoWR. It solves the NLTE radiative 
transfer in a spherically expanding atmosphere simultaneously with the statistical-equilibrium 
equations and accounts at the same time for energy conservation. 
Iron-group line blanking is treated by means of the superlevel approach 
\citep{graefeneretal2002}, and a wind clumping in first-order approximation is taken 
into account \citep{hamannetal2004}. We do not calculate hydrodynamically consistent 
models, but assume a velocity field following a $\beta$-law with $\beta\,=\,1$.
For the PoWR models, we adopted the elemental abundances that we found in our 
TMAP analysis. We extrapolated the mass and the luminosity of \sao using the evolutionary 
tracks from \citet{Hall2013} and found $M=$0.36\,\Msol\,\,and $\log$($L$/\Lsol)$=$2.5. We also 
calculated models with $M=$0.47\,\Msol\,\,and $\log$($L$/\Lsol)$=$2.2 (values extrapolated from the 
tracks by \citealt{driebe1998} ), but we found that the line profiles change only slightly with 
$M$ and $L$. This agrees with \citet{heraldbianchi2007} who report that wind features 
are not very sensitive to the gravity, which is in the case of the PoWR code an equivalent input parameter.

Based on these assumptions, we varied the terminal wind velocity $v_\infty$, mass-loss rate \Mdot, 
and \Teff\ in our models. The relative strengths of \ion{N}{V}\,/\,\ion{C}{IV} turned out to be 
very sensitive to \Teff, so that we can achieve an error of only $\Delta$\Teffw{\pm3}. The sensitvity to \Teff\ can 
be clearly seen by comparing the best-fit models for 1988 and 1992, where the only difference of these models 
is \Teff. In the \Teffw{38} model for 1988, \ion{C}{IV} is relatively stronger than \ion{N}{V}, whereas the opposite 
holds for the \Teffw{43} model for 1992. The mass-loss rate was derived from the strengths of the P-Cygni profiles. 
As the P-Cygni profiles broaden as $v_\infty$ increases, $v_\infty$ can be measured not only from the blue edge of 
the absorption component of the P-Cygni profiles, but also from the shape of the emission peak.

\begin{figure}
  \resizebox{\hsize}{!}{\includegraphics{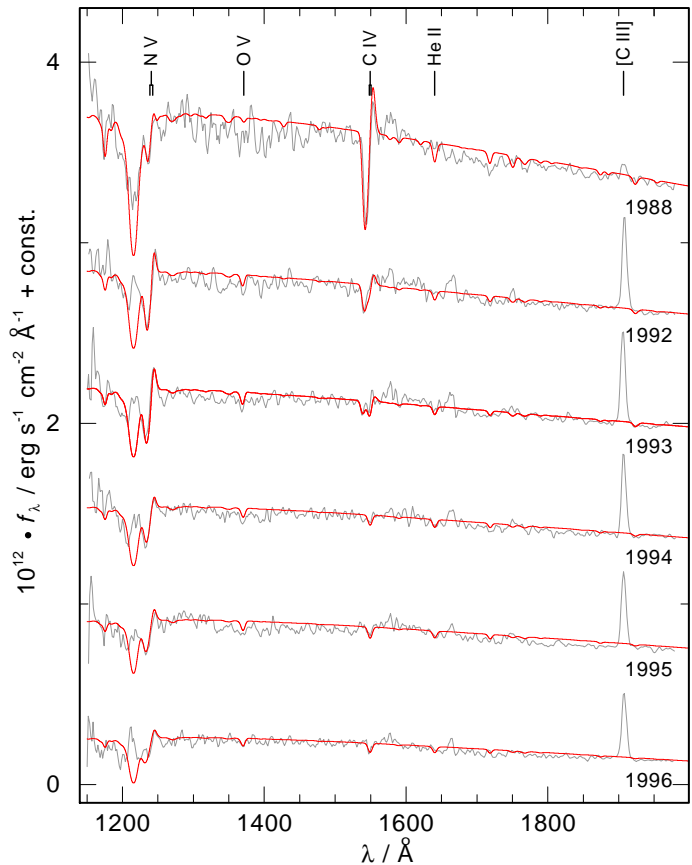}}
  \caption{Comparison of the IUE spectra taken from 1988 to 1996 (thin, gray) with the best-fit PoWR model (thick red). All but the 1988 observation are shifted for clarity.} 
  \label{fig:wind}
\end{figure}

\begin{figure}
  \resizebox{\hsize}{!}{\includegraphics{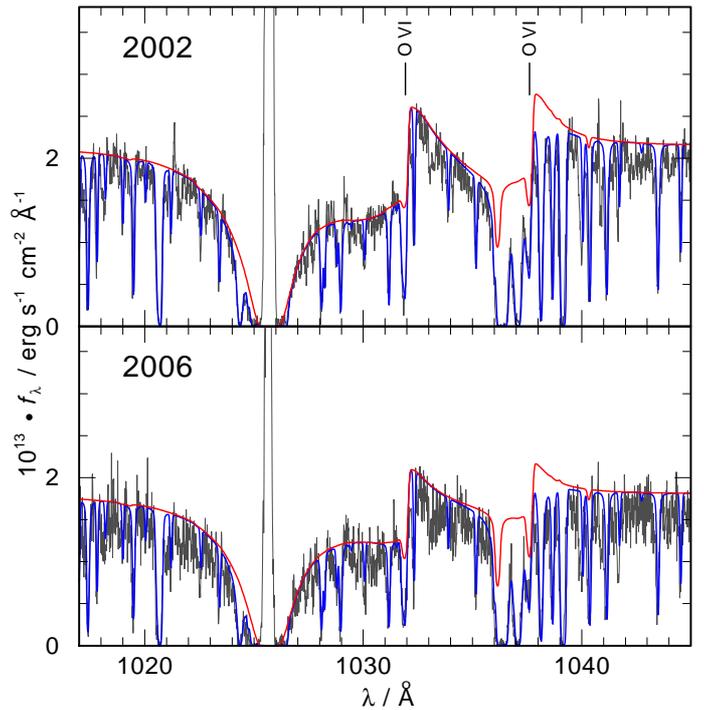}}
  \caption{Comparison the FUSE spectra taken in 2002 and 2006 (gray) with the best-fit PoWR model (thick red). The blue line (thin) indicates the combined stellar and ISM spectrum.} 
  \label{fig:wind2}
\end{figure}

The best-fit models are shown in Figs.\,\ref{fig:wind} (IUE spectra) 
and \ref{fig:wind2} (FUSE). The P-Cygni profile of \Ionw{O}{6}{1037.6} 
in the PoWR models is blended by interstellar H$_2$ and is therefore not visible in the observation. 
Furthermore, we found that the absorption components of \Ionww{O}{6}{1031.9, 1037.6} must have an 
interstellar origin. Table\,\ref{tab:wind} lists the parameters that we found from this analysis. The 
resulting surface gravities are also given. We point out that these values refer to the models with 
$M=$0.36\,\Msol, the resulting values for the surface gravity from the $M=$0.47\,\Msol\,\,models were 
about 0.1\,--\,0.2 dex higher. 

We find that the mass-loss rate decreased continuously from $\log$(\Mdot\,/\,\Msol\,yr$^{-1})=\,-9.0$ in 1988 to 
$\log$(\Mdot\,/\,\Msol\,yr$^{-1})=\,-11.6$ in 2006.
The previously published value ($v_\infty = 3500$\,km/\,s, \citealt{partha1995}) turned out to be an overestimate. We 
find that the terminal wind velocity has steadily increased from $v_\infty = 1800$\,km/\,s in 1988 to 
$v_\infty = 2800$\,km/\,s in 2006.

\section{Results and discussion} 
\label{sect:discussion}

We analyzed the high-resolution UV spectra taken with FUSE, STIS, 
FOS, and IUE by means of non-LTE line blanketed model atmospheres. The values of \Teff\ were derived with 
high precision by evaluating the ionization equilibria of 
\ion{C}{III}\,/\,\ion{C}{IV} (FOS, STIS and FUSE spectra), 
\ion{O}{III}\,/\,\ion{O}{IV},  
\ion{S}{IV}\,/\,\ion{S}{V}\,/\,\ion{S}{VI} (FUSE spectra), 
and \ion{N}{III}\,/\,\ion{N}{IV} (FUSE 2002 spectra). From the IUE spectra, we 
derived \Teff\ using the relative strength of 
\ion{N}{V}\,/\,\ion{C}{IV}.\\
Both the surface gravity and the H/He ratio (solar) were obtained by an analysis of the 
\Ion{He}{2} lines found in the STIS and FOS spectra. For the FOS observations, the best 
fit was found with \loggw{5.5}, while for the STIS observation the model with \loggw{6.0} 
fits better. This value of \loggw{6.0} was confirmed by the wings of the \Ion{He}{2} lines 
in the FUSE spectra.

\sao exhibits solar abundances of H, He, O, and N. The C abundance 
is definitely subsolar. This indicates that the AGB phase of the star
was terminated before the third dredge-up \citep{Mello2012}. P and S were found to be subsolar, the Si and Fe 
abundances are solar. For Ne, Mg, and Ni, we could only derive upper limits 
(five times solar for Ne, ten times solar for Na, solar for the rest). We could not find any hint of a change 
in the chemical abundances during the years. Within the error limits (typically $\pm$0.3\,dex) the abundances found for the CS are 
in good agreement with the nebula abundances (Table ~\ref{tab:neb}) determined by \citet{Arkhipova2013} and \citet{partha1993}.

\begin{table}
	\centering
	\caption{Comparison of the element abundances (number fractions relatively to H) 
                 of \sao (CS) as derived by our analysis and of its PN as derived by analysis of the spectrum in 1990 \citep{partha1993}, 1992, and 2011 \citep{Arkhipova2013}.}
        \label{tab:neb}
        \setlength{\tabcolsep}{.5em}
        \begin{tabular}{r@{\,/\,} l r@{.} l r@{.} l r@{.} l r@{.} l}
	\hline
	\hline
         \multicolumn{2}{l}{}    & \multicolumn{2}{c}{CS}  &  \multicolumn{2}{l}{}    &  \multicolumn{2}{l}{PN}    &  \multicolumn{2}{l}{}\\
         \multicolumn{2}{l}{}    & \multicolumn{2}{l}{}    &  \multicolumn{2}{l}{2011}  &  \multicolumn{2}{l}{1992}  &  \multicolumn{2}{l}{1990}\\
        \hline
        \noalign{\smallskip}
        He&H   &   8&50$\times 10^{-2}$   & 9&60$\times 10^{-2}$ & 9&30$\times 10^{-2}$ &  1&03$\times 10^{-1}$  \\  
        C&H    &   5&69$\times 10^{-5}$   & \multicolumn{2}{c}{} & 7&59$\times 10^{-5}$ &  \multicolumn{2}{c}{}  \\  
        N&H    &   6&69$\times 10^{-5}$   & 5&75$\times 10^{-5}$ & 6&45$\times 10^{-5}$ &  6&50$\times 10^{-5}$   \\  
        O&H    &   4&84$\times 10^{-4}$   & 1&91$\times 10^{-4}$ & 2&46$\times 10^{-4}$ &  3&00$\times 10^{-4}$   \\  
        Ne&H &$<$\,3&59$\times 10^{-4}$   & 3&47$\times 10^{-5}$ & 5&75$\times 10^{-5}$ &  9&20$\times 10^{-5}$   \\  
        S&H    &   6&03$\times 10^{-6}$   & 2&40$\times 10^{-6}$ & 6&03$\times 10^{-6}$ &  2&20$\times 10^{-6}$   \\  
   	\hline 
	\end{tabular}
\end{table}

\begin{figure}
  \resizebox{\hsize}{!}{\includegraphics{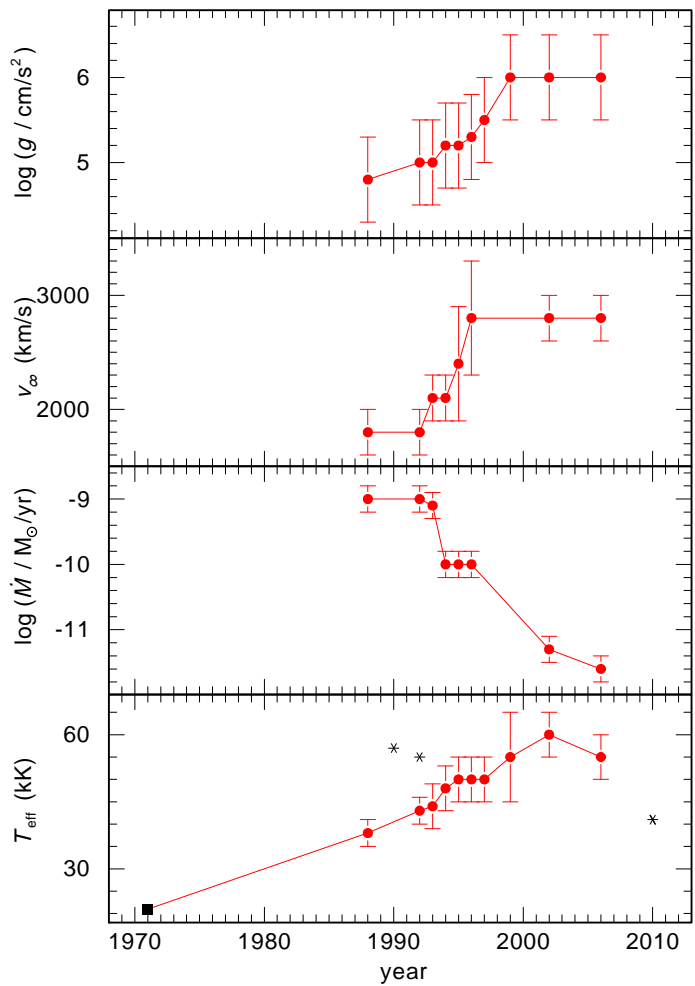}}
  \caption{Temporal evolution of \Teff, \logg, mass-loss rate \Mdot, and the terminal wind velocities $v_\infty$ as derived in our analysis (red). The black stars in the bottom panel show the \Teff\ as derived by \citet{Arkhipova2013}, the black square indicates the \Teff\ estimate when \sao was still a B-type supergiant (1971).} 
  \label{fig:temp}
\end{figure}

The temporal evolution of the atmospheric parameters of \sao is summarized in Table ~\ref{tab:wind} 
and Fig.~\ref{fig:temp}. We found that compared to the temperature found from the spectrum in 1971 
(\se{sect:introduction}) \sao has increased its \Teff\ by about 40\,kK within only thirty 
years. The peak \Teff\ of 60\,kK was reached in 2002, and the 2006
observations suggest that it is now decreasing. 
We cannot confirm the values for \Teff\ found by \citet{Arkhipova2013}. They find the peak \Teff\ 
in 1990 (\Teffw{57}), i.e. about 10\,kK higher than our analysis from spectra taken at that time. We want to 
point out, that the formalism of \citet{Kaler1978} used by \citet{Arkhipova2013} might not provide good estimates 
for the CS temperature. Comparing the \Teff\ of \citet{Kaler1978} found for his sample of CSPNe, shows often 
big deviations from recent literature values (e.g. \object{NGC\,6543}: \citet{Kaler1978} found 
\Teffw{45} while \citet{heraldbianchi2011} found \Teffw{60}). 
However, evaluation of the ionization equilibria found in the FUSE spectra shows \Teff\ declining from 2002 on, 
supporting the contention by \citet{Arkhipova2013} that \Teff\ has recently peaked and has begun to decline.

By examining the P-Cygni profiles of the \ion{C}{IV} and \ion{N}{V} 
resonance doublets in the IUE spectra from 1988 to 1996 and the P-Cygni profiles of 
\ion{O}{VI} in the FUSE spectra, we derived the mass-loss rates and terminal wind 
velocities as a function of time. We found a steady decrease in the mass-loss rate 
(from $\log$(\Mdot\,/\,\Msol\,yr$^{-1})=\,-9.0$ in 1988 to $\log$(\Mdot\,/\,\Msol\,yr$^{-1})=\,-11.6$ 
in 2006) and the steady increase in the terminal wind velocity (from $v_\infty = 1800$\,km/\,s 
in 1988 to $v_\infty = 2800$\,km/\,s in 2006). 
To compare the mass-loss rate with predictions from radiative-driven wind theory, we used  
Eq. (5) in \citet{VinkCassisi2002} with $L=2.5$\,\Lsol, $M=0.354$\,\Msol, \Teffw{38}, 
and $Z=2$ to derive the theoretical value of the mass-loss rate in 1988. We found 
$\log \dot{M} = -8.8\pm0.3$, which is in good agreement with the observed value. However, we 
note that the formula given by \citet{VinkCassisi2002} might not give the right predictions, 
since it is meant for a slightly different parameter space in $L$, \Teff, and $M$. 

The decrease in flux with a simultaneous increase of \Teff\ from 1988 to 1999 can be explained by the 
contraction of the star ($L \propto R^2 \Teff^4 \propto g^{-1} \Teff^4$). This is consistent with the 
increase in \logg\ as seen in the FOS (1997) and STIS (1999) observations, and consistent with the increase 
in $v_\infty$ which is proportional to $v_{\mathrm{esc}} \propto g$. 
The decrease in flux from 2002 to 2006 can be explained by a decrease in \Teff\ as seen 
in the FUSE spectra. 

\subsection{Distance}
\label{sect:distance}

The distance of \sao was estimated first by \citet{kozok1985} using UBV photometry and the assumption of a 
typical absolute magnitude of a Be star, leading to $d=5.64$\,kpc. \citet{Arkhipova2013} argues that 
\citet{kozok1985} overestimated the absolute magnitude of \sao since it was not a main sequence star at 
that time. Using the estimates of \citet{partha1993} for \Teffw{37.5} and \loggw{4.0}, they estimated 
$M \approx 0.55$\,\Msol\ and derived $d = 1.8$\,kpc. 
We calculated the spectroscopic distance of \sao using the flux calibration 
of \citet{heberetal1984} for $\lambda_\mathrm{eff} = 5454\,\mathrm{\AA}$,
$$d[\mathrm{pc}]=7.11 \times 10^{4} \cdot \sqrt{H_\nu\cdot M \times 10^{0.4\, m_{\mathrm{V}_0}-\log g}} \,\, ,$$
\noindent
with $m_\mathrm{V_o} = m_\mathrm{V} - 2.175 c$, $c = 1.47 E_\mathrm{B-V}$, and 
the Eddington flux $H_\nu$ ($7.15\times 10^{-4}$ erg/cm$^{2}$/s/Hz) at 5454\,\AA\,\,of our model atmosphere 
(\Teffw{50} and \loggw{5.5}). The visual brightness ($m_{\mathrm{V}}=14.9$) was extracted from the FOS spectrum, 
so we used all the model parameters found for this year. We assumed  $E_\mathrm{B-V}=0.14$ (\se{sect:ism}) and 
a stellar mass of $M=0.354^{+0.14}_{-0.05}$\,\Msol\ (\se{sect:mass}). 
We derived $d=1.6^{+0.8}_{-1.2}$\,kpc which leads to a height above the Galactic plane of $z=0.3 \pm 0.2$\,kpc. These 
values are in good agreement with the estimates of \citet{Arkhipova2013}. From the angular diameter 
of 2$\farcs$3, a linear radius of the PN of $R=0.009^{+0.004}_{-0.007}$\,pc results. With the expansion 
velocity of 8.4\,km/s \citep{Arkhipova2013}, the kinematic age of the PN is only about $1013^{+488}_{-793}$\,years. 
We note that different masses, which we derive from various evolutionary scenarios, hardly affect the distance determination 
and that the large errors are mainly due to the uncertainties in \logg.

\subsection{Mass and evolutionary status}
\label{sect:mass}

\begin{figure}
  \resizebox{\hsize}{!}{\includegraphics{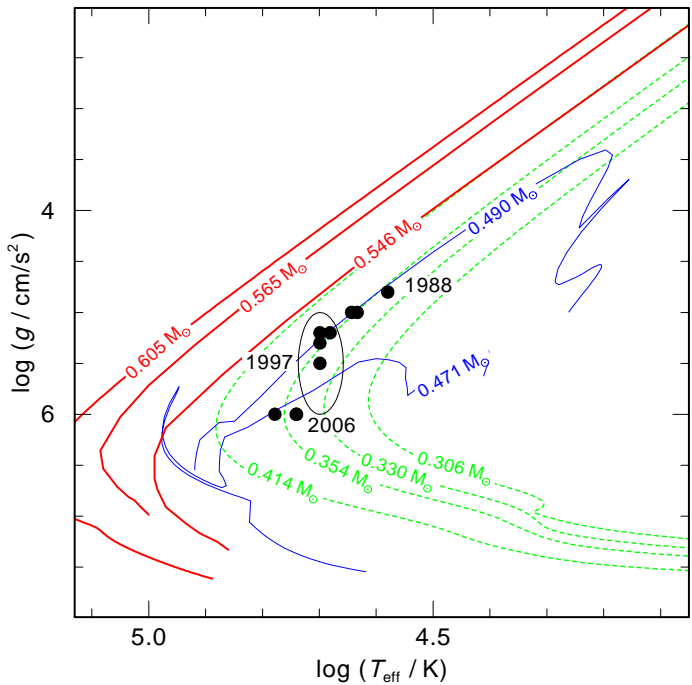}}
  \caption{Evolution of \sao (black dots) in the $\log$ \Teff\,-- \logg plane compared to 
           post-AGB (red, thick) by \citet{bloecker1995},
           post-EHB (blue, thin) by \citet{dorman1993}, and 
           post-RGB (green, dashed) evolutionary tracks by \citet{Hall2013} The tracks are labeled with stellar masses. 
           The ellipse indicates the errors of \Teff\ and \logg\ in 1997.} 
  \label{fig:postAGB}
\end{figure}

In Fig.~\ref{fig:postAGB}, we compare the time-dependent location of \sao 
in the $\log$ \Teff\,-- \logg plane with H-rich post-AGB evolutionary tracks by \citet{bloecker1995}, 
post-EHB-tracks by \citet{dorman1993}, and post-RGB-tracks by \citet{Hall2013}. 
The last represent the evolution of post-common envelope remnants of 
post-RGB stars that will not ignite He and become low-mass white dwarfs with a He core. 
The determination of the mass of \sao would only be possible, if the evolutionary status of this 
object is known. For that reason, we give only the mass estimates valid for 
a certain evolutionary scenario below.

The position of \sao in the log \Teff\ -- \logg\ plane places the star in the region of sdO stars.
Since \sao does not match the 0.546\,\Msol\,\,track within the error ranges, it may either be a very 
low-mass post-AGB star with a mass of about 0.53\,\Msol\,\,or even an AGB-manqu\'{e} star, which never 
reaches the AGB.

The existence of the PN around \sao suggests that it is more likely a post-AGB star 
than an AGB-manqu\'{e} star because --
in general -- only these stars are expected to eject a PN. If we consider \sao as a 0.53\,\Msol\,\,
post-AGB star, it evolves much more quickly than predicted by theory. Comparing \Teffw{21} estimated for 
1971 to the peak value of \Teffw{60} in 2002, \sao increased its temperature to 40\,kK within only 30 years. 
According to Fig\@.\,6 in \citet{bloecker1995} such a rapid evolution ($\log (dt/dT_\mathrm{eff}/\mathrm{yrK}^{-1}) = -3$) 
would be expected only for a 0.87\Msol\,\,star with an initial mass of about 6\,\Msol. For a 0.55\Msol\,\,star it would 
take 130000 years to heat up from \Teffw{21} to \Teffw{60}.\\
An alternative to the canonical post-AGB evolution is a late thermal pulse (LTP,  e.g. \citealt{Bloecker2001}). 
The evolutionary speed of objects, which are considered to have undergone a (very) late He-shell flash 
(V605 Aql, e.g., \citealt{Clayton2006}, V4334 Sgr, e.g., \citealt{Hajduk2005} and FG Sge, e.g., \citealt{Jeffery2006}), 
is very high (decades). A very late thermal pulse can be ruled out because this scenario produces a 
hydrogen-free stellar surface, whereas a normal surface composition is typical of a LTP in the phase relevant to \sao. 
The LTP occurs when the star evolves with roughly constant luminosity from the AGB towards the white dwarf domain. 
The convective shell triggered by excessive helium burning is not able to penetrate the hydrogen-rich envelope from 
below because the entropy jump across the helium/hydrogen interface is too large. Only when the star evolves back to its 
Hayashi limit on the AGB (\Teff$\lesssim$\,7000\,K), envelope convection sets in again 
\citep{BloeckerSchoenberner1996, BloeckerSchoenberner1997, Schoenberner2008}. Therefore, the surface abundances of \sao 
are in agree with the LTP scenario. The evolutionary calculations of \cite{Bloecker2001} for a 0.625\,\Msol\,\,LTP star predict 
that the solar surface composition of the star holds up after the LTP. Only when the star has cooled down to \Teffw{10}, the H abundance 
decreases while the He and C abundances increase. The LTP scenario is able to explain the rapid evolution (increase in \Teff\, and 
drop in luminosity). It would predict a decrease in \Teff\ (the FUSE observation in 2006 already suggests that) and an increase in 
the brightness within the next decades (similar to the evolution of FG Sge). The LTP is a good candidate for explaining the 
evolutionary status of \sao; however, concrete evolutionary models that match the position of \sao in the $\log$ \Teff\,-- \logg 
plane, are missing to prove this scenario.\\

Considering \sao as an AGB-manqu\'{e} star, which ignited central He-burning, we derive a mass of 
0.49\,\Msol\,\,according to the tracks of \citet{dorman1993}. Within this scenario the existence 
of the PN and the short evolutionary time scales are even more difficult to explain. Even if the RGB precursor 
would have ejected a PN, it should have disappeared a long time ago since the horizontal branch (HB) evolution 
lasts $\approx 10^8$ years. It is worthwhile mentioning that the mass of the remaining H layer 
($\le$ 0.001\,\Msol) of extended horizontal branch (EHB) stars is much too low to produce 
a nebula at the end of the HB stage. We can also rule out an early hot-flasher scenario. \citet{Brown2001} 
found that as the Reimers mass-loss parameter $\eta_{\mathrm{R}}$ increases, the peak of the main helium core 
flash shifts to higher temperatures, and the subsequent zero age horizontal branch (ZAHB) position becomes hotter. 
However their models never produced stars with a ZAHB position hotter than \Teffw{31.5}.

Recently \citet{Hall2013} have proposed that it is possible that at least some PNe 
are composed of matter ejected from a binary star system during common envelope (CE) 
evolution. For these PNe, the ionizing component is the hot and luminous remnant of 
a giant that had its envelope ejected by a companion in the process of spiraling in to its current 
short-period orbit. A large fraction of CE phases that end with ejection of the envelope are thought 
to be initiated by low-mass red giants, giants with inert, degenerate helium cores. In their 
calculations \citet{Hall2013} find, that PNe are expected in post-CE systems with core
masses greater than about 0.3\,\Msol\,\,if remnants end the CE phase in thermal equilibrium. 
Considering this scenario for \sao, we estimate a mass of $0.354\pm0.03$\,\Msol. The rapid 
evolution of \sao, though, is also not predicted in their models. A higher mass-loss rate might 
increase the evolutionary speed, but there are no hints of any fenhanced mass loss in the observations. 
However, if the remnant is in thermal non-equilibrium after the CE ejection, it might evolve rapidly 
enough (Philip Hall, priv. comm.). 

The current decrease in \Teff\ of \sao from 2002 on found in our work and also by 
\citet{Arkhipova2013} argues for post-RGB evolution instead of the post-EHB evolution, since the 
H-shell-burning models cool down directly, whereas the He-core burning post-EHB models become even hotter 
(Fig.~\ref{fig:postAGB}).

The bipolar shape of the PN is most likely due to some kind of binary interaction 
of a RGB or AGB star with either a stellar or substellar companion. This was already suggested by 
\citet{Bobrowsky1994} to explain the axisymmetric structure of \sao. \citet{soker1997} proposed that bipolar 
structures of PNe, as seen in \sao, can be explained by close binary interactions that avoid 
a CE phase, or they entered this phase only in their late evolution. The 
gravitational interaction between the companion and the PN progenitor and/or their winds are 
non-negligible. The companion's tidal force can spin up the primary and enhance the mass-loss 
rate or the companion can accrete mass from the primary's wind. This accretion process is likely 
to result in a high mass-loss rate in the equatorial plane, leading to a bipolar PN. 
The possible companion detected by \citet{Bobrowsky1998}, however, is too far away 
(638\,AU at a distance of 1.6\,kpc) for significant interaction with \sao. But it might be possible 
that \sao has another, hitherto undetected companion.

\subsection{Comparison with other low-mass CSPNe}
\label{sect:comparison}

\sao is not the only low-mass CSPN. \citet{napi1999} found five such candidates 
(\object{HDW\,11}, 
 \object{K\,2$-$2}, 
 \object{GD\,561}, 
 \object{PHL\,932}, and 
 \object{DeHt\,5}) that are not in accordance with post-AGB evolution. 
For the DA-type WD \object{DeHt\,5} ($M\,=\,$0.414\Msol), they considered a post-RGB evolution as the 
kinematical age of the PN ($\approx$129\,000 yrs) is in good agreement with the post-RGB age 
($\approx$100\,000\,yrs) estimated from the calculations of \citet{driebe1998}. 
The two DAO-type WDs \object{K\,2$-$2} ($M\,=\,$0.39\Msol) and \object{HDW\,1}1 ($M\,=\,$0.38\Msol) 
are also candidates for this evolutionary scenario.

For \object{PHL\,932} and \object{EGB\,5}, an object with similar parameters, the post-RGB times 
corresponding to the tracks of \citet{driebe1998} are much longer than the ages of the PNe. The PN 
\object{EGB\,5} is believed to be a remnant of an ejected CE. \citet{Geier2011} discovered a close 
low-mass companion orbiting the sdB central star and thus \object{EGB\,5} is considered to be a good 
candidate for a post-CE object. 
\citet{napi1999} considered \object{PHL\,932} ($M\,=\,$0.28\Msol) to 
be the outcome of a CE event of a star with a degenerate CO core as calculated by \citet{IbenTutukov1985} 
or even as the result of a merger within a common envelope \citep{mendezetal1988b}. However \citet{Frew2010} 
convincingly demonstrate that the nebula around \object{PHL\,932} is not a PN, but rather a 
Str\"omgren sphere in the ambient ISM. Also \object{EGB\,5} has recently been classified as an ionized \Ion{H}{2} 
region \citep{Frew2013}. 
The same holds for the nebulae around the DAO-type WDs GD\,561 
and BD-22$^{\circ}$3467 \citep{ziegler2012}, which \citet{FrewParker2010} also consider to be 
Str\"omgren spheres. Other objects of this type are the sdO stars 
\object{HD\,497898}, 
\object{LSS\,630}, 
\object{LSE\,44},
\object{LSE\,153}, 
\object{SB\,705}, and 
\object{KPD\,0720$-$0003} around 
which nebulosities were detected that are not believed to be ejected by or physically associated with 
the corresponding stars \citep{mendezetal1988a}. This can clearly not be true for \sao owing to its 
complex nebula structure and the expansion velocity.\\
A multi-shell planetary nebula around the hot sdO star \object{2MASS J19310888+4324577} 
has recently been detected \citep{alleretal2013}. The morphology of the nebula (bipolar and elliptical shell, 
whose major axes are oriented perpendicular to each other) strongly resembles the one of the Stingray Nebula. 
The complex nebula structure and the fact that \object{2MASS J19310888+4324577} was found to be 
in a binary system \citep{jacoby2012} suggest that it might have formed through binary star evolution.\\

\section{Conclusions}
\label{sect:conclusions}

\sao is a rapidly evolving object. Its evolutionary status remains unclear.  
The most reasonable explanations are a late He-shell flash or CE evolution with a 
remnant that is in thermal non-equilibrium after the CE ejection. However, respective models are lacking that match the 
position of \sao in the $\log$ \Teff\,-- \logg plane.
The contradiction between observations and theory make \sao particularly 
interesting. Its fast evolution gives us the unique opportunity to study stellar evolution in real 
time and establishes constraints for stellar evolutionary theory.
Further observations, in the next years, decades and even centuries, are essential for monitoring whether the rapid evolution of 
\sao is still going on and to see if it is directly evolving to the white dwarf domain or back to the AGB. The 
detection of a close binary would support the scenario of a CE ejection, whereas an increase in brightness and 
decrease in \Teff\ over the next decades, would indicate an evolution back to the AGB and hence speak for a LTP scenario.

\begin{acknowledgements}
NR is supported by the German Research Foundation (DFG, grant WE 1312/41-1),
TR by the German Aerospace Center (DLR, grant 05\,OR\,0806). 
We thank Marcelo Miguel Miller Bertolami and Philip Hall for helpful discussions and comments. 
MP is grateful to Profs\@. Ajit K\@. Kembhavi, Kandaswamy Subramanian 
and T\@. Padmanabhan for their kind encouragement, support, and hospitality.
This research has made use of the SIMBAD database, operated at the CDS, Strasbourg, France.
This research made use of NASA's Astrophysics Data System.
This work used the profile-fitting procedure OWENS developed by M\@. Lemoine 
and the French FUSE Team.
Some of the data presented in this paper were obtained from the 
Mikulski Archive for Space Telescopes (MAST). STScI is operated by the 
Association of Universities for Research in Astronomy, Inc., under NASA 
contract NAS5-26555. Support for MAST for non-HST data is provided by 
the NASA Office of Space Science via grant NNX13AC07G and by other grants 
and contracts.
\end{acknowledgements}

\bibliographystyle{aa}
\bibliography{AA_SAO} 

\end{document}